

\def\leaderdot{\leaders\hbox to 1 em {\hss.\hss}\hfill}

\dimen0= \parindent                           
\dimen1= \hsize \advance\dimen1 by -\dimen0

\dimen2=\baselineskip
\def\skiplines#1 { \dimen3=\dimen2 \multiply\dimen3 by #1 \vskip \dimen3}

\def\fullline{\hbox to \fullhsize}

\def\numpage{\baselineskip=24pt\fullline{\the\footline}}

\def\mathcedilla{\vtop{\hbox{c}{\kern0pt\nointerlineskip}
	         {\hbox{$\mkern-2mu \mathchar"0018\mkern-2mu$}}}}

\mathchardef\gq="0060
\mathchardef\dq="0027
\mathchardef\ssmath="19
\mathchardef\aemath="1A
\mathchardef\oemath="1B
\mathchardef\omath="1C
\mathchardef\AEmath="1D
\mathchardef\OEmath="1E
\mathchardef\Omath="1F
\mathchardef\imath="10 
\mathchardef\fmath="0166
\mathchardef\gmath="0167
\mathchardef\vmath="0176




\def\charlvmidlw#1#2{\,\vtop{\ialign{##\crcr
      #1\crcr\noalign{\kern1pt\nointerlineskip}
      $\hfil#2\hfil$\crcr}}\,}

\def\charlvlowlw#1#2{\,\vtop{\ialign{##\crcr
      $\hfil#1\hfil$\crcr\noalign{\kern1pt\nointerlineskip}
      #2\crcr}}\,}

\def\charlvmidup#1#2{\,\vbox{\ialign{##\crcr
      $\hfil#1\hfil$\crcr\noalign{\kern1pt\nointerlineskip}
      #2\crcr}}\,}

\def\charlvupup#1#2{\,\vbox{\ialign{##\crcr
      #1\crcr\noalign{\kern1pt\nointerlineskip}
      $\hfil#2\hfil$\crcr}}\,}

\def\emptybox{\vbox{\kern.7ex\hbox{\kern.5em}\kern.7ex}}

\font\tenmib=cmmib10
\newfam\bfmitfam

\textfont\bfmitfam=\tenmib
\scriptfont\bfmitfam=\seveni
\scriptscriptfont\bfmitfam=\fivei


\def\twodot{.\kern-0.1em.}

\def\paral{\mathrel{/\kern-.25em/}}
\def\grlo{\mathrel{\hbox{\lower.2ex\hbox{\rlap{$>$}\raise1ex\hbox{$<$}}}}}
\def\logr{\mathrel{\hbox{\lower.2ex\hbox{\rlap{$<$}\raise1ex\hbox{$>$}}}}}
\def\greq{\mathrel{\hbox{\lower1ex\hbox{\rlap{$=$}\raise1.2ex\hbox{$>$}}}}}
\def\loeq{\mathrel{\hbox{\lower1ex\hbox{\rlap{$=$}\raise1.2ex\hbox{$<$}}}}}
\def\grsim{\mathrel{\hbox{\lower1ex\hbox{\rlap{$\sim$}\raise1ex\hbox{$>$}}}}}
\def\losim{\mathrel{\hbox{\lower1ex\hbox{\rlap{$\sim$}\raise1ex\hbox{$<$}}}}}

\def\uniset{\rlap{\ninerm 1}\kern.15em 1}

\def\emptysq{\mathbin{\vbox{\hrule\hbox{\vrule height1ex \kern.5em 
                            \vrule height1ex}\hrule}}}
\def\emptyrect{\mathbin{\vbox{\hrule\hbox{\vrule height1ex \kern1em 
                              \vrule height1ex}\hrule}}}
\def\rightonleftarrow{\mathrel{\hbox{\raise.5ex\hbox{$\rightarrow$}\ignorespaces
                                   \lower.5ex\hbox{\llap{$\leftarrow$}}}}}
\def\leftonrightarrow{\mathrel{\hbox{\raise.5ex\hbox{$\leftarrow$}\ignorespaces
                                   \lower.5ex\hbox{\llap{$\rightarrow$}}}}}
\def\BB{{\rm I\kern-.17em B}}
\def\BC{{\rm \kern.24em
            \vrule width.05em height1.4ex depth-.05ex
            \kern-.26em C}}
\def\BD{{\rm I\kern-.17em D}}
\def\BE{{\rm I\kern-.17em E}}
\def\BF{{\rm I\kern-.17em F}}
\def\BG{{\rm \kern.24em
            \vrule width.05em height1.4ex depth-.05ex
            \kern-.26em G}}
\def\BH{{\rm I\kern-.22em H}}
\def\BI{{\rm I\kern-.22em I}}
\def\BJ{{\rm \kern.19em
            \vrule width.02em height1.5ex depth0ex
            \kern-.20em J}}
\def\BK{{\rm I\kern-.22em K}}
\def\BL{{\rm I\kern-.17em L}}
\def\BM{{\rm I\kern-.22em M}}
\def\BN{{\rm I\kern-.20em N}}
\def\BO{{\rm \kern.24em
            \vrule width.05em height1.4ex depth-.05ex
            \kern-.26em O}}
\def\BP{{\rm I\kern-.17em P}}
\def\BQ{{\rm \kern.24em
            \vrule width.05em height1.4ex depth-.05ex
            \kern-.26em Q}}
\def\BR{{\rm I\kern-.17em R}}
\def\BT{{\rm \kern.24em
            \vrule width.02em height1.5ex depth 0ex
            \kern-.27em T}}
\def\BU{{\rm \kern.30em
            \vrule width.02em height1.47ex depth-.05ex
            \kern-.32em U}}
\def\BZ{{\rm Z\kern-.32em Z}}


\def\CB{{\cal B}}

\def\CE{{\cal E}}

\def\CN{{\cal N}}

\def\CR{{\cal R}}

\def\CU{{\cal U}}



\def\sl{\it}


\def\Temp{\the\catcode`\@}
\catcode`\@=11
\ifx\TypeSizes@Loaded\relax
  \message{TypeSizes already loaded}
  \catcode`\@=\Temp
    \else \let\TypeSizes@Loaded=\relax\catcode`\@=\Temp\fi

\newfam\roundfam

\font\fiveround=cmsy5 scaled \magstephalf

\font\sixrm=cmr6
\font\sixi=cmmi6
\font\sixsy=cmsy6
\font\sixbf=cmbx6
\font\sixround=cmsy6 scaled \magstephalf

\font\sevenrm=cmr7
\font\seveni=cmmi7
\font\sevensy=cmsy7
\font\sevenbf=cmbx7
\font\sevenit=cmti7
\font\sevenround=cmsy7 scaled \magstephalf

\font\eightrm=cmr8
\font\eighti=cmmi8
\font\eightsy=cmsy8
\font\eightbf=cmbx8
\font\eighttt=cmtt8
\font\eightit=cmti8
\font\eightsl=cmsl8
\font\eightround=cmsy8 scaled \magstephalf

\font\ninerm=cmr9
\font\ninei=cmmi9
\font\ninesy=cmsy9
\font\ninebf=cmbx9
\font\ninett=cmtt9
\font\nineit=cmti9
\font\ninesl=cmsl9
\font\nineround=cmsy9 scaled \magstephalf

%
%
\font\tenround=cmsy10 scaled \magstephalf
\font\sc=cmcsc10

\font\twelverm=cmr10 scaled \magstep1
\font\twelvei=cmmi10 scaled \magstep1
\font\twelvesy=cmsy10 scaled \magstep1
\font\twelvebf=cmbx10 scaled \magstep1
\font\twelvett=cmtt10 scaled \magstep1
\font\twelveit=cmti10 scaled \magstep1
\font\twelvesl=cmsl10 scaled \magstep1
\font\twelveround=cmsy10 scaled 1314 

%
\catcode`@=11 
\newskip\ttglue

\def\sixpoint{\def\rm{\fam0\sixrm}
  \textfont0=\sixrm \scriptfont0=\fiverm \scriptscriptfont0=\fiverm
  \textfont1=\sixi  \scriptfont1=\fivei  \scriptscriptfont1=\fivei
  \textfont2=\sixsy \scriptfont2=\fivesy \scriptscriptfont2=\fivesy
  \textfont3=\tenex   \scriptfont3=\tenex \scriptscriptfont3=\tenex
  \textfont\itfam=\sevenit  \def\it{\fam\itfam\sevenit}%
  \textfont\slfam=\eightsl  \def\sl{\fam\slfam\eightsl}%
  \textfont\ttfam=\eighttt  \def\tt{\fam\ttfam\eighttt}%
  \textfont\bffam=\sixbf    \scriptfont\bffam=\sixbf
   \scriptscriptfont\bffam=\fivebf  \def\bf{\fam\bffam\sixbf}%
\textfont\roundfam=\sixround  \def\round{\fam\roundfam\sixround}%
\scriptfont\roundfam=\fiveround \scriptscriptfont\roundfam=\fiveround
  \tt \ttglue=.5em plus .25em minus .15em
  \normalbaselineskip=7.2pt
  \normallineskip=0.6pt
  \setbox\strutbox=\hbox{\vrule height5.1pt depth2.1pt width0pt}%
  \def\strut{\relax\ifmmode\copy\strutbox\else\unhcopy\strutbox\fi}%
  \let\sc=\fiverm  \let\big=\sixbig  \normalbaselines\rm}

\def\sevenpoint{\def\rm{\fam0\sevenrm}
  \textfont0=\sevenrm \scriptfont0=\sixrm \scriptscriptfont0=\fiverm
  \textfont1=\seveni  \scriptfont1=\sixi  \scriptscriptfont1=\fivei
  \textfont2=\sevensy \scriptfont2=\sixsy \scriptscriptfont2=\fivesy
  \textfont3=\tenex   \scriptfont3=\tenex \scriptscriptfont3=\tenex
  \textfont\itfam=\sevenit  \def\it{\fam\itfam\sevenit}%
  \textfont\slfam=\eightsl  \def\sl{\fam\slfam\eightsl}%
  \textfont\ttfam=\eighttt  \def\tt{\fam\ttfam\eighttt}%
  \textfont\bffam=\sevenbf  \scriptfont\bffam=\sixbf
   \scriptscriptfont\bffam=\fivebf  \def\bf{\fam\bffam\sevenbf}%
\textfont\roundfam=\sevenround  \def\round{\fam\roundfam\sevenround}%
\scriptfont\roundfam=\sixround \scriptscriptfont\roundfam=\fiveround
  \tt \ttglue=.5em plus .25em minus .15em
  \normalbaselineskip=8.4pt
  \normallineskip=0.7pt
  \setbox\strutbox=\hbox{\vrule height5.95pt depth2.45pt width0pt}%
  \def\strut{\relax\ifmmode\copy\strutbox\else\unhcopy\strutbox\fi}%
  \let\sc=\sixrm  \let\big=\sevenbig  \normalbaselines\rm}

\def\eightpoint{\def\rm{\fam0\eightrm}
  \textfont0=\eightrm \scriptfont0=\sixrm \scriptscriptfont0=\fiverm
  \textfont1=\eighti  \scriptfont1=\sixi  \scriptscriptfont1=\fivei
  \textfont2=\eightsy \scriptfont2=\sixsy \scriptscriptfont2=\fivesy
  \textfont3=\tenex   \scriptfont3=\tenex \scriptscriptfont3=\tenex
  \textfont\itfam=\eightit  \def\it{\fam\itfam\eightit}%
  \textfont\slfam=\eightsl  \def\sl{\fam\slfam\eightsl}%
  \textfont\ttfam=\eighttt  \def\tt{\fam\ttfam\eighttt}%
  \textfont\bffam=\eightbf  \scriptfont\bffam=\sixbf
   \scriptscriptfont\bffam=\fivebf  \def\bf{\fam\bffam\eightbf}%
\textfont\roundfam=\eightround  \def\round{\fam\roundfam\eightround}%
\scriptfont\roundfam=\sixround \scriptscriptfont\roundfam=\fiveround
  \tt \ttglue=.5em plus .25em minus .15em
  \normalbaselineskip=9.6pt
  \normallineskip=0.8pt
  \setbox\strutbox=\hbox{\vrule height6.8pt depth2.8pt width0pt}%
  \def\strut{\relax\ifmmode\copy\strutbox\else\unhcopy\strutbox\fi}%
  \let\sc=\sixrm  \let\big=\eightbig  \normalbaselines\rm}

\def\ninepoint{\def\rm{\fam0\ninerm}
  \textfont0=\ninerm \scriptfont0=\sixrm \scriptscriptfont0=\fiverm
  \textfont1=\ninei  \scriptfont1=\sixi  \scriptscriptfont1=\fivei
  \textfont2=\ninesy \scriptfont2=\sixsy \scriptscriptfont2=\fivesy
  \textfont3=\tenex   \scriptfont3=\tenex \scriptscriptfont3=\tenex
  \textfont\itfam=\nineit  \def\it{\fam\itfam\nineit}%
  \textfont\slfam=\ninesl  \def\sl{\fam\slfam\ninesl}%
  \textfont\ttfam=\ninett  \def\tt{\fam\ttfam\ninett}%
  \textfont\bffam=\ninebf  \scriptfont\bffam=\sixbf
   \scriptscriptfont\bffam=\fivebf  \def\bf{\fam\bffam\ninebf}%
\textfont\roundfam=\nineround  \def\round{\fam\roundfam\nineround}%
\scriptfont\roundfam=\sixround \scriptscriptfont\roundfam=\fiveround
  \tt \ttglue=.5em plus .25em minus .15em
  \normalbaselineskip=10.8pt
  \normallineskip=0.9pt
  \setbox\strutbox=\hbox{\vrule height7.65pt depth3.15pt width0pt}%
  \let\sc=\sevenrm  \let\big=\ninebig  \normalbaselines\rm}

\def\tenpoint{\def\rm{\fam0\tenrm}
  \textfont0=\tenrm \scriptfont0=\sevenrm \scriptscriptfont0=\fiverm
  \textfont1=\teni  \scriptfont1=\seveni  \scriptscriptfont1=\fivei
  \textfont2=\tensy \scriptfont2=\sevensy \scriptscriptfont2=\fivesy
  \textfont3=\tenex   \scriptfont3=\tenex \scriptscriptfont3=\tenex
  \textfont\itfam=\tenit  \def\it{\fam\itfam\tenit}%
  \textfont\slfam=\tensl  \def\sl{\fam\slfam\tensl}%
  \textfont\ttfam=\tentt  \def\tt{\fam\ttfam\tentt}%
  \textfont\bffam=\tenbf  \scriptfont\bffam=\sevenbf
   \scriptscriptfont\bffam=\fivebf  \def\bf{\fam\bffam\tenbf}%
\textfont\roundfam=\tenround  \def\round{\fam\roundfam\tenround}%
\scriptfont\roundfam=\sevenround \scriptscriptfont\roundfam=\fiveround
  \tt \ttglue=.5em plus .25em minus .15em
  \normalbaselineskip=12pt
  \normallineskip=1pt
  \setbox\strutbox=\hbox{\vrule height8.5pt depth3.5pt width0pt}%
  \let\sc=\eightrm  \let\big=\tenbig  \normalbaselines\rm}

\def\twelvepoint{\def\rm{\fam0\twelverm}
  \textfont0=\twelverm \scriptfont0=\tenrm \scriptscriptfont0=\eightrm
  \textfont1=\twelvei  \scriptfont1=\teni  \scriptscriptfont1=\eighti
  \textfont2=\twelvesy \scriptfont2=\tensy \scriptscriptfont2=\eightsy
  \textfont3=\tenex   \scriptfont3=\tenex \scriptscriptfont3=\tenex
  \textfont\itfam=\twelveit  \def\it{\fam\itfam\twelveit}%
  \textfont\slfam=\twelvesl  \def\sl{\fam\slfam\twelvesl}%
  \textfont\ttfam=\twelvett  \def\tt{\fam\ttfam\twelvett}%
  \textfont\bffam=\twelvebf  \scriptfont\bffam=\tenbf
   \scriptscriptfont\bffam=\eightbf  \def\bf{\fam\bffam\twelvebf}%
\textfont\roundfam=\twelveround  \def\round{\fam\roundfam\twelveround}%
\scriptfont\roundfam=\tenround \scriptscriptfont\roundfam=\eightround
  \tt \ttglue=.5em plus .25em minus .15em
  \normalbaselineskip=14.4pt
  \normallineskip=1.2pt
  \setbox\strutbox=\hbox{\vrule height10.2pt depth4.2pt width0pt}%
  \let\big=\fourteenbig  \normalbaselines\rm}






\def\dist{distribution}

\def\etal{{\it et al.}}







\def\NP{Neyman-Pearson}






\def\UPMC{Universit\'e Pierre et Marie Curie}


\def\JASA{{\it Journal of the American Statistical Association} }

\input psfig.sty
\magnification=\magstep1
\baselineskip=14pt

\def\NP{{\CN}_+}
\def\NPM{{\CN}_-^+}
\def\mub{{\mu^-}}
\def\muu{{\mu^+}}

\def\bga{{\bf \gamma}}
\def\bmu{{\bf \mu}}
\def\bte{{\bf \theta}}
\def\bV{{\bf V}}
\def\enn{^{(n)}}
\def\enm{^{(n-1)}}
\def\rho{\varrho}

\hrule height 0pt

\vglue .2 true in
\centerline{\bf Simulation of truncated normal variables}

\vglue .2 true in
\centerline{Christian P.\ Robert}
\centerline{\it LSTA, \UPMC, Paris}

\vglue .375 true in
\centerline{\bf Abstract}
\vglue .2 true in

We provide in this paper simulation algorithms for one-sided and
two-sided truncated normal distributions. These algorithms are then
used to simulate multivariate normal variables with restricted
parameter space for any covariance structure. 



\baselineskip=15pt
\parskip=2pt

\vskip 1truecm
\centerline{\bf 1.\ \ Introduction}
\medskip

The need for simulation of truncated normal variables 
appears in Bayesian inference
for some truncated parameter space problems.
Indeed, it is rarely the case that analytical computations
are possible and numerical integration can be very intricated for
large dimensions. Typical examples of such setups can be found in
order restricted (or {\it isotonic}) regression, as illustrated in
Robertson, Wright and Dykstra (1988). 
For instance, one can consider a $n\times n$
table of normal random variables $x_{ij}$ with means $\theta_{ij}$ which
are increasing in $i$ and $j$ $(1\le i,j\le n)$, as in Dykstra and
Robertson (1982). When $n$ is large, both maximum likelihood and 
Bayesian inferences on this table can be quite cumbersome and 
simulation techniques are then
necessary to either obtain mle's by {\it stochastic restoration} 
(see Qian and
Titterington, 1991) or Bayes estimators by {\it Gibbs sampling} 
(see Gelfand and
Smith, 1990). Gibbs sampling actually provides a large set of examples where
simulation from truncated distributions is necessary, for instance
for {\it censored models} since the recovery 
of the censored observations implies
simulation from the corresponding truncated distribution, as shown in 
details by Gelfand, Smith and Lee (1992). See also Chen and
Deely (1992) who propose a new version of the Gibbs sampler for estimating
the ordered coefficients of a regression model.

\medskip
We first construct in Section 2 an efficient algorithm for unidimensional
truncated normal variables. This algorithm is quite simple and, in the
particular case of one-sided truncated normal distributions, it slightly
improves on a previous algorithm developed by Marsaglia (1964).
Our multidimensional extension in Section 3  is also based
on this algorithm. Actually, we propose to use Gibbs sampling
to reduce the simulation problem to a sequence of
one-dimensional simulations. The resulting sample, being derived from a 
Markov chain, is not independent, but can be used similarly for all
estimation purposes.

\vskip .25truein
\centerline{\bf 2.\ \ The univariate case}
\medskip

\noindent {\bf 2.1. One-sided truncation.} \ 
Let us denote $\NP(\mu,\mub,\sigma^2)$ the truncated normal distribution
with {\it left} truncation point $\mub$, i.e.\ the distribution with density
$$
f(x|\mu,\mub,\sigma^2) = {\exp(-(x-\mu)^2/2\sigma^2) \over
\sqrt{2\pi}\sigma (1-\Phi((\mub-\mu)/\sigma)) }\,\BI_{x\ge \mub} .
$$
Obviously, a readily available method is to simulate from a normal
distribution $\CN(\mu,\sigma^2)$ until the generated number is larger
than $\mub$. This method is quite reasonable when $\mub<\mu$ but is of
no use when $\mub$ is several standard deviations to the right of $\mu$.
Similarly, Gelfand \etal\ (1992) and Chen and Deely (1992) suggest to
use the classical c.d.f.\ inversion technique, namely to simulate
$u\sim\CU_{[0,1]}$ and to take
$$
z = \mu + \Phi^{-1} \left( 
\Phi\left( {\mub - \mu \over \sigma} \right) + u \left\{ 1-
\Phi\left( {\mub - \mu \over \sigma} \right) \right\} \right)
$$
as the simulation output, 
but this method calls for a simultaneous evaluation of the
normal c.d.f.\ $\Phi$ and of its inverse $\Phi^{-1}$, and may be quite
inefficient if $\mub-\mu$ is large, since the precision of the approximation
of $\Phi$ then strongly matters.
We provide below an {\it accept-reject} algorithm which is more efficient than
repeatedly simulating from the normal distribution as soon as $\mub>\mu$.
In the sequel, we will assume without loss of generality 
that $\mu=0$ and $\sigma^2=1$,
since the usual location-scale rescaling allows to standardize truncated
normal variables.

\medskip
Let us recall first that the general accept-reject algorithm is based
on the following result (see Devroye, 1985, pp.\ 40-60).

\medskip
{\narrower\noindent
{\bf Lemma 2.1} \ Let $h$ and $g$ be two densities such that 
$h(x) \le Mg(x)$ for every $x$ in the support of $h$.
The random variable $x$ resulting from the following algorithm 

\item{\bf 1.} {\it Generate $z\sim g(z)$;} 

\item{\bf 2.} {\it Generate $u\sim\CU_{[0,1]}$. If $u\le h(z)/Mg(z)$, 
take $x=z$; otherwise, repeat from step 1.}

\smallskip\noindent
is distributed accorded to $h$.

\medskip}

In our case, a possible choice for $g$ is the translated exponential
distribution $\CE xp(\alpha,\mub)$ with density
$$
g(z|\alpha,\mub) = \alpha e^{-\alpha(z-\mub)} \ \BI_{z\ge \mub}.
$$
Since, for $z\ge \mub$, we have
$$
e^{\alpha(z-\mub)} e^{-z^2/2} \le e^{\alpha^2/2 -\mub \alpha}
$$
if $\alpha>\mub$ and
$$
e^{\alpha(z-\mub)} e^{-z^2/2} \le e^{-(\mub)^2/2}
$$
if $\alpha\le\mub$, the constant $M$ is given by
$$
\cases{
  {\alpha \over \sqrt{2 \pi}(1-\Phi(\mub))} e^{\alpha^2/2 - \alpha\mub}
		&if $\alpha\ge \mub$, \cr
  {\alpha \over \sqrt{2 \pi}(1-\Phi(\mub))} e^{-(\mub)^2/2}
		& otherwise \cr
}
$$ 
and the ratio $h(z)/M g(z)$ by
$$
{h(z)\over M g(z)} = \cases{
e^{-z^2/2 + \alpha(z-\mub) - \alpha^2/2 + \alpha \mub}
		&if $\alpha\ge \mub$, \cr
e^{-z^2/2 + \alpha(z-\mub) +(\mub)^2/2}
		& otherwise. \cr
}
$$
We then derive from Lemma 2.1 the corresponding accept-reject algorithm.

\medskip
{\narrower\noindent
{\bf Lemma 2.2} \ The following algorithm

\item{\bf 1.} {\it Generate $z\sim\CE xp(\alpha,\mub)$;}

\item{\bf 2.} {\it Compute $\rho(z) = \exp(-(\alpha-z)^2/2)$ 
if $\mub<\alpha$
and $\rho(z) = \exp((\mub-\alpha)^2/2) \allowbreak
\exp(-(\alpha-z)^2/2)$ otherwise;}

\item{\bf 3.} {\it Generate $u\sim\CU_{[0,1]}$ and take $x=z$ 
if $u\le\rho(z)$; otherwise, repeat from step 1.}

\smallskip\noindent
leads to the generation of a random variable from $\NP(0,\mub,1)$.

\medskip}
Now, noticing that the probability of acceptance in one single run is
$$
\BE_\alpha[\rho(z)] = \cases{
\alpha e^{\alpha\mub - \alpha^2/2} \Phi(-\mub) \sqrt{2\pi}
&if $\mub<\alpha$, \cr
\alpha e^{(\mub)^2/2} \Phi(-\mub) \sqrt{2\pi}
&otherwise,
}
$$
we deduce that the optimal scale factor in the exponential distribution
attained for 
$$
\alpha^*(\mub) = {\mub+\sqrt{(\mub)^2+4} \over 2}
$$ 
in the first case and for $\alpha=\mub\,$ in
the second case. Furthermore, since the corresponding probabilities are
proportional to 
$$
\alpha^*(\mub) e^{\mub \alpha^*(\mub)/2} / \sqrt{e}
$$
and $\,\mub \exp((\mub)^2/2)\,$ respectively, 
with the same coefficient of proportionality, it can be shown by using
the reparametrization in $\alpha^*$ (i.e.\ $\mub=\alpha^*-1/\alpha^*$)
that the first probability is always
greater and that the best choice of $\alpha$ is $\alpha^*(\mub)$.
Therefore,

\medskip
{\narrower\noindent
{\bf Proposition 2.3} \ The optimal exponential accept-reject algorithm
to simulate from a $\NP(0,\mub,1)$ when $\mub>0$ is given by

\smallskip
\item{\bf 1.} {\it Generate $z\sim \CE xp(\alpha^*,\mub)$;}

\item{\bf 2.} {\it Compute $\rho(z) = \exp\{-(z-\alpha^*)^2/2\}$;}

\item{\bf 3.} {\it Generate $u\sim\CU_{[0,1]}$ and take $x=z$
if $u\le\rho(z)$; otherwise, go back to step 1.}

\medskip}

Table 2.1 below gives the expected probability $\BE_{\alpha^*}[\rho(z)]$ 
for several values of $\mub$. It shows the gain brought by using this
accept-reject algorithm since the probability of accepting in one
passage is $0.760$ for $\mub=0$, as compared with $0.5$ for the
repeated normal sampling alternative. The improvement increases as
$\mub$ goes away from $0$ and the probability of accepting goes to $1$
as $\mub$ goes to infinity. Note that the
probability of accepting is greater than
$$
\mub e^{(\mub)^2/2} \Phi(-\mub) \sqrt{2\pi},
$$
probability of accepting for $\alpha=\mub$; this is also 
the rate obtained by Marsaglia (1964) when proposing an accept-reject
algorithm using the tail of a Raleigh distribution (see
also Devroye, 1985, pp. 380-382). The improvement brought by using
$\CE xp(\alpha^*,\mub)$ is significant for the moderate values of $\mub$. Those
large probabilities also hint at likely improvements over repeated
normal sampling even when $\mub<0$, but such developments would call 
for much more elaborated algorithms and, moreover, fast normal generators
can overcome the advantages of using a more complex algorithm.

\medskip
$$
\matrix{ \mub &0 &0.5 &1 &1.5 &2 &2.5 &3 \cr
& & & & & & & \cr
\BE_{\alpha^*}[\rho(z)] 
&$\quad$0.760 &0.826 &0.876 &0.910 &0.934 &0.950 &0.961 \cr
}
$$
\smallskip
\centerline{{\bf Table 2.1 - }  Average probability of acceptance}
\centerline{ according to the truncation point $\mub$.}

\medskip
Simulation from the {\it right truncated normal distribution}, 
$x\sim\CN_-(\mu,\muu,\sigma^2)$, can be directly derived from the above
algorithm since $-x\sim\CN_+(-\mu,-\muu,\sigma^2)$. We consider in the
next section the simulation from the {\it two-sided truncated normal
distribution} for which modifications of the above algorithm are necessary.

\bigskip \noindent
{\bf 2.2. Two-sided truncated normal distribution.} \ When considering
the two-sided truncated normal distribution $\NPM(\mu,\mub,\muu,\sigma^2)$,
with density
$$
f(x|\mu,\mub,\muu,\sigma) = {e^{-(x-\mu)^2/2\sigma^2} \over
\sqrt{2\pi} \sigma\, [\Phi((\muu-\mu)/\sigma))-\Phi(\mub-\mu)/\sigma))]} ,
$$
the simulation method heavily depends on the range $\muu-\mub$. As before,
a first possibility is to simulate from a $\CN(\mu,\sigma^2)$ distribution
until $z\in[\mub,\muu]$ (or even to invert the c.d.f.). 
However, if $\Phi(\muu-\mu)-\Phi(\mub-\mu)$ is
small or even if $(\mub-\mu)(\muu-\mu)>0$, more efficient alternatives are
available. We propose here to consider, in addition to the previous algorithms,
an accept-reject approach based on the uniform \ $\CU_{[\mub,\muu]}$ 
distribution.
Once again, we can assume without loss of generality that $\mu=0$ and 
$\sigma^2=1$.

\medskip
The accept-reject algorithm based on $\CU_{[\mub,\muu]}$ is

{\narrower
\item{\bf 1.} {\it Generate $z\sim\CU_{[\mub,\muu]}$;} 

\item{\bf 2.} {\it Compute}
$$
\rho(z)= \cases{ \exp(-z^2/2) &{\it if} \ \ $0\in[\mub,\muu]$ \cr
\exp(\{(\muu)^2-z^2\}/2) &{\it if} \ \ $\muu<0$ \cr
\exp(\{(\mub)^2-z^2\}/2) &{\it if} \ \ $0<\mub$ \cr
}
$$

\item{\bf 3.} {\it Generate $u\sim\CU_{[0,1]}$ and take $x=z$ if $u\le\rho(z)$;
otherwise, go back to step 1.}

\smallskip}

\noindent
The corresponding expected probability of running the above algorithm only
once is
$$
\eqalignno{
\BE[\rho(z)] &= \int_\mub^\muu e^{-z^2/2} dz\ {e^d\over {\muu-\mub} } \cr
	  &= \sqrt{2\pi}\ {e^d\over \muu-\mub}\ (\Phi(\muu)-\Phi(\mub) ) \cr
}
$$
where $d=0,\ (\muu)^2/2$ or $(\mub)^2/2$\ whether $\muu\mub<0,\ \muu<0$ or
$\mub>0$. Therefore, when $\muu\mub<0$, it is more efficient to use this
algorithm rather than to use
the repeated normal method if $\muu-\mub<\sqrt{2\pi}$.

\medskip
We now oppose simulation from the uniform algorithm to repeated simulation from
a one-sided truncated normal distribution. For instance, if $\mub>0$, we
simulate $z\sim\NP(0,\mub,1)$ until $z<\muu$. Using the optimal algorithm
of Proposition 2.3, the probability of accepting in one passage is
$$
\eqalign{
P(u\le\rho(z) &\hbox { and } z\le \muu) \cr
	&= \int_\mub^\muu e^{-(z-\alpha^*)^2/2} 
       \alpha^*	e^{-\alpha^* (z-\mub)} dz 
						\cr
&= \alpha^* e^{ \alpha^*\mub-(\alpha^*)^2/2 } \sqrt{2\pi}
		\,(\Phi(\muu)-\Phi(\mub)) 
							\cr
&= \alpha^* e^{ \alpha^*\mub/2 } \sqrt{2\pi/e}
		\,(\Phi(\muu)-\Phi(\mub))\, .  \cr
}
$$
Therefore, it is better to use the truncated $\NP(0,\mub,1)$ algorithm if
$$
\alpha^* e^{\alpha^* \mub /2 } /\sqrt{e} > {e^{\mub^2/2} \over \muu-\mub}
$$
i.e.\ if
$$
\muu > \mub + {2\sqrt{e}\over \mub+\sqrt{\mub^2+4} }
\exp\left\{ { \mub^2 - \mub \sqrt{\mub^2+4} \over 4 } \right\}.
\eqno(2.1)
$$

\centerline{\psfig{file=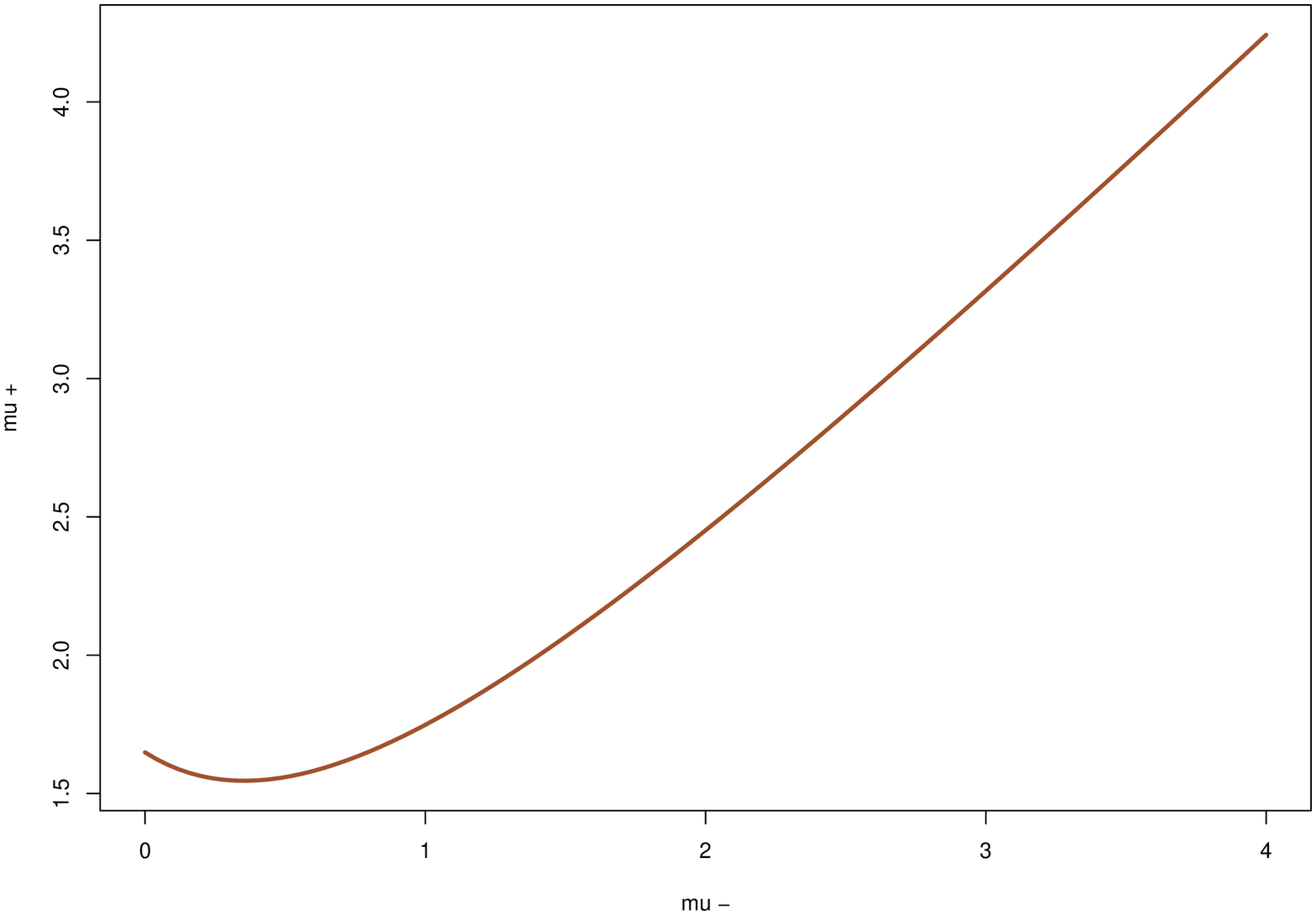,height=7truecm,angle=270}}

\centerline{{\bf Figure 2.1 - }  Lower bound (2.1) on $\muu$}
\centerline{for the use of the truncated normal algorithm.}

Figure 2.1.\ provides the lower bound of (2.1) as a function of $\mub$.
Note that, as $\mub$ increases, the range $\muu-\mub$ has to get smaller
for uniform accept-reject sampling to be used.
The corresponding decomposition is straightforward to derive when $\muu<0$.
Table 2.2 below gives the expected probabilities of acceptance in one run 
for several values of $\mub$ and $\muu-\mub$.

$$
\matrix{ \cr \cr \muu-\mub\cr}
\qquad \matrix{
 &  &  & \mub &    & \cr
 &  &  & 	&    & \cr
 & 0 &0.5 &1 & 1.5 & 2 \cr
2 & \ \ .726 \ & .811 & .869 & .907 & .932 \cr
1 & .856 & .687 & .751 & .826 & .878 \cr
0.5 & .960 & .851 & .759 & .680 & .679 \cr
0.1 & .998 & .974 & .950 & .927 & .905 \cr}
$$
\smallskip
\centerline{{\bf Table 2.2 - } Average probabilities of acceptance}
\centerline{for the simulation of $\NPM(0,\mub,\muu,1)$.}

\vskip .25truein
\centerline{\bf 3.\ \ The multivariate case}
\medskip

We consider now a multivariate normal distribution $\,\CN_p(\bmu,\Sigma)$ 
restricted to a convex subset $\CR$ of $\BR^p$, denoted
$\CN^T(\bmu,\Sigma,\CR)$. We assume that the
one-dimensional slices of $\CR$,
$$
\CR_i(\theta_1,\ldots,\theta_{i-1},\theta_{i+1},\ldots,\theta_p) = 
\left\{ \theta_i;  (\theta_1,\ldots,\theta_{i-1},
\theta_i,\theta_{i+1},\ldots,\theta_p) \in \CR \right\},
$$
are readily available, in the sense that these sets can be represented
as intervals $[\theta^-_i,\theta^+_i]$, where the bounding functions 
$\theta^-_i$
and $\theta_i^+$, depending on
$(\theta_1,\ldots,\theta_{i-1},\theta_{i+1},\ldots,\theta_p)$,
are easily computable $(1\le i\le p)$.

\medskip
The algorithm we propose below belongs to the class of {\it Markov Chain
Monte-Carlo methods\/} (as referred to in Hastings (1970) and Geyer (1991)).
Namely, instead of generating a sequence $\bte_k$ of i.i.d.\ random
vectors from the distribution of interest, we provide a
sequence $\bte\enn$ which is a Markov chain with stationary distribution
the distribution of interest. Such an approximation may seem to fall far
from the mark but results like the {\it ergodic theorem} ensure that the
average of any quantity of interest $f(\bte)$,
$$
{1\over N}\sum_{n=1}^N f(\bte\enn),
\eqno(3.1)
$$
is converging to the expectation  $\BE[f(\bte)]$ as $N$ goes to infinity, 
thus generalizing the law of large numbers. More details on the
application of Markov chain theory in this setup are given in 
Ripley (1987, pp.\ 113-114), Geyer (1991)
and Tierney (1991). Following the early {\it Metropolis algorithm}
(Metropolis \etal, 1953), Markov
chain Monte-Carlo simulation methods have been used extensively in the
past years in {\it Gibbs sampling} theory for Bayesian computation
(see Tanner and Wong (1987), Gelfand and Smith (1990) and Tanner (1991)).
The main difficulty of this approach, as opposed to
usual (independent) Monte-Carlo methods, is to monitor the convergence
of the chain  to the stationary distribution. Apart from classical
central limit theorem (see Geyer, 1991) and
time-series methods (see Ripley, 1987, chap.\ 6), one can suggest the
simultaneous estimation of several quantities until approximate 
stationarity of the corresponding averages (3.1) is attained for all
functions. Gelman and Rubin (1991) also suggest to run several
times the algorithm with drastically different starting values.
In our particular setup, convergence to the stationary \dist\ should
be particularly fast since the compactness of $\CR$ ensures
{\it geometric convergence} (see Tierney, 1991).

\medskip
In the setup of truncated normal distributions, the Markov chain
$\bte\enn$ is obtained by generating successively the components 
of $\CN^T(\bmu,\Sigma,\CR)$, i.e.
$$
\eqalign{
&{\bf 1.}\ \theta_1\enn \sim \NPM(\BE[\theta_1|\theta_2\enm,\ldots,
\theta_p\enm],\theta^-_1,\theta_1^+,\sigma^2_1) \cr
&{\bf 2.}\ \theta_2\enn \sim \NPM(\BE[\theta_2|\theta_1\enn,\theta_3\enm,
\ldots, \theta_p\enm],\theta_2^-,\theta^+_2,\sigma^2_2) \cr
&\qquad \qquad \ldots \cr
&{\bf p.}\ \theta_p\enn \sim \NPM(\BE[\theta_p|\theta_1\enn,\ldots,
\theta_{p-1}\enn],\theta_p^-,\theta^+_p,\sigma^2_p) \cr
}
$$
where the expectations and variances in the above truncated normal
distributions are the conditional 
(non-truncated) expectations and variances of the
$\theta_i$ given $\theta_{\neg i}=
(\theta_1,\ldots,\theta_{i-1},\allowbreak\theta_{i+1},\ldots,\theta_p)$. Namely,
we have
$$
\eqalign{
\BE[\theta_i|\theta_{\neg i}] &= \mu_i + \Sigma^t_{i\neg i}
\Sigma^{-1}_{\neg i\neg i} (\theta_{\neg i} -\mu_{\neg i} ), \cr
\sigma^2_i &= \sigma^2_{ii} - 
\Sigma^t_{i\neg i} \Sigma^{-1}_{\neg i\neg i}
\Sigma_{i\neg i}, \cr
}
$$
where $\Sigma_{\neg i\neg i}$ is the $(p-1)\times(p-1)$ matrix 
derived from $\Sigma=(\sigma^2_{ij})$ by
eliminating its $i$-th row and its $i$-th column and $\Sigma_{i\neg i}$
is the $(p-1)$ vector derived from the 
$i$-th column of $\Sigma$ by removing the $i$-th row term. 

\medskip
Moreover, it is important to note that
there is no need to invert all the matrices
$\Sigma_{\neg i\neg i}$ 
to run the algorithm. Indeed, it is possible to
derive these inverses from the global inverse matrix
$\bV=\Sigma^{-1}$ since they can be written
$$
\Sigma^{-1}_{\neg i\neg i} 
= \bV_{\neg i\neg i} - \bV_{i\neg i}\bV_{i\neg i}^t/\bV_{ii},
\eqno(3.2)
$$
where $\bV_{\neg i\neg i}$ and $\bV_{i\neg i}$ are derived from $\bV$
the way $\Sigma_{\neg i\neg i}$ and $\Sigma_{i\neg i}$ are derived from
$\Sigma$.  Therefore, the algorithm 
only requires at most one inversion of $\Sigma$ and 
the computation of the submatrices $\Sigma^{-1}_{\neg i\neg i}$ by (3.2).

\medskip
The comparison with a classical rejection-sampling method based on the
simulation of $x\sim\CN_p(\mu,\Sigma)$ until the result belongs to $\CR$
is quite delicate, depending on the probability $P(x\in \CR)$ but also
on the overall purpose of the simulation. In fact, if this probability is
rather large and a single observation from $\CN^T(\mu,\Sigma,\CR)$ is needed,
it is clear that rejection sampling is preferable. On the contrary, if a large
sample is needed, as it is the case for Gibbs sampling and related maximum
likelihood methods, then the Markov chain Monte-Carlo method should be
superior, especially if $\CR$ is small, since as mentioned above, convergence
of the Gibbs sampler to the stationary \dist\ should be fast.

\medskip
As a concluding remark, let us consider the following example. The truncated
distribution of interest is
$$
\left( \matrix{\theta_1\cr \theta_2\cr} \right) \sim
\CN^T\left( {\bf 0}, \left[\matrix{1 &\rho \cr \rho & 1\cr}\right],\CR\right),
$$
with truncation space $\CR$ the ball $\CB(\bga,r)$ of center $\bga=(\gamma_1,
\gamma_2)$ and radius $r$. Therefore,
$$
\eqalign{
\theta^-_1(\theta_2) = \gamma_1 - \sqrt{r^2 - (\gamma_2-\theta_2)^2}, \qquad
&\theta^+_1(\theta_2) = \gamma_1 + \sqrt{r^2 - (\gamma_2-\theta_2)^2}, \cr
\theta^-_2(\theta_1) = \gamma_2 - \sqrt{r^2 - (\gamma_1-\theta_1)^2}, \qquad
&\theta^+_2(\theta_1) = \gamma_2 + \sqrt{r^2 - (\gamma_1-\theta_1)^2} \cr
}
$$
and the conditional distributions defining the Markov chain are
$$
\eqalign{
{\bf 1.}\ \ &\theta_1\enn\sim\NPM\left(\rho\theta_2\enm,
\theta^-_1(\theta_2\enm),
\theta^+_1(\theta_2\enm),1-\rho^2\right) \cr
{\bf 2.}\ \ &\theta_2\enn\sim\NPM\left(\rho\theta_1\enn,
\theta^-_2(\theta_1\enn),
\theta^+_2(\theta_1\enn),1-\rho^2\right). \cr
}
$$

\vskip .25truein
\noindent
{\bf Acknowledgements}

This research was performed while
visiting Cornell University. The author is grateful to
George Casella for his support through NSF Grant No.\ DMS9100839 and
NSA Grant No.\ 90F-073
and to Charles McCulloch for
pointing out the single inversion argument in the multivariate
case and helpful comments. By mentioning a mistake in an earlier version,
Ranjini Natarajan also led to an improvement in the efficiency of the
algorithms.

\bigskip
\parskip=2pt \baselineskip=15pt

\centerline{\bf References}
\medskip
{\advance\leftskip by 0.2in \parindent=-0.2in

Casella, G.\ and George, E.I.\ (1991)
Explaining the Gibbs sampler.  {\it The Amer.\ Statist.} (to appear).

Chen, M.H.\ and Deely, J. (1992) Application of a new Gibbs
Hit-and-Run sampler to a constrained linear multiple regression
problem. Tech.\ report, Purdue University, Lafayette, IN.

Devroye, L. (1985) {\it Non-Uniform Random Variate Generation}.
Springer-Verlag, New York.

Dykstra, R.L.\ and Robertson, T. (1982) An algorithm for isotonic regression
for two or more independent variables. {\it Ann. Statist.} {\bf 10}, 
708-716.

Gelfand, A.E. and Smith, A.F.M. (1990)
Sampling based approaches to calculating
marg-\break
inal densities.  {\it JASA} {\bf 85}, 398--409.

Gelfand, A.E., Smith, A.F.M.\ and Lee, T.M.\ (1992) Bayesian analysis
of constrained parameter and truncated data problems using Gibbs sampling.
{\it JASA} {\bf 87}, 523-532.

Gelman, A.\ and Rubin, D.B.\ (1991) A single series from the Gibbs
sampler provides a false sense of security. In {\it Bayesian Statistics 4},
J.O.\ Berger, J.M.\ Bernardo, A.P.\ Dawid and A.F.M.\ Smith (Eds.). Oxford
University Press.

Geyer, C.J.\ (1991) Markov Chain Monte Carlo Maximum Likelihood. To
appear in {\it Computer Sciences and Statistics: Proc. 23d Symp. Interface}.

Hastings, W.K. (1971) Monte-Carlo sampling methods using Markov chains and 
their applications. {\it Biometrika} {\bf 57}, 97-109.

Marsaglia, G. (1964) Generating a variable from the tail of a normal
distribution. {\it Technometrics} {\bf 6}, 101-102.

Metropolis, N., Rosenbluth, A.W., Rosenbluth, M.N., Teller, A.H.\ and
Teller, E. (1953) Equations of state calculations by fast computing machines.
{\it J.\ Chemical Phys.} {\bf 21}, 1087-1091.

Qian, W.\ and Titterington, D.M.\ (1991) Estimation of parameters in hidden
Markov models. {\it Phil. Trans. Royal Soc. London\/} A {\bf 337}, 407-428. 

Ripley, B.D. (1987) {\it Stochastic simulation}. J. Wiley, New York.

Robertson, T., Wright, F.T.\ and Dykstra, R.L. (1988) {\it Order
Restricted Statistical Inference}. J. Wiley, New York.

Tanner, M. (1991) {\it Tools for Statistical Inference}.
Springer-Verlag, New York.

Tanner, M. and Wong, W. (1987) The calculation of posterior
distributions by data augmentation. \JASA\ {\bf 82}, 528-550.

Tierney, L. (1991) Markov chains for exploring posterior distributions. To
appear in {\it Computer Sciences and Statistics: Proc. 23d Symp. Interface}.

}

\bigskip\noindent
{LSTA, Bo\^\i te 158}\hfill{\it March 1992}
\leftline{Universit\'e Paris 6}
\leftline{4, place Jussieu}
\leftline{75252 Paris Cedex 5 - France}

\bye